\documentclass[prd,amsmath,amssymb,superscriptaddress,12pt]{revtex4-1}

\usepackage{color, colortbl}
\usepackage{graphicx}
\usepackage{dcolumn}
\usepackage{mathtools}
\usepackage{bm}
\newcommand{\beq}{\begin{equation}}
\newcommand{\eeq}{\end{equation}}
\newcommand{\ba}{\begin{array}{ccc}}
\newcommand{\ea}{\end{array}}

\newcommand{\ketbra}[2]{\left|#1\middle\rangle\middle\langle#2\right|}
\newcommand{\ket}[1]{\left| #1 \right\rangle}
\newcommand{\bra}[1]{\left\langle #1 \right|}
\def\bea{\begin{eqnarray}}
\def\eea{\end{eqnarray}}
\usepackage{color}
\usepackage{float}
\usepackage{url}

\begin{document}
\title{Reconstructing quantum states with generative models}
\author{Juan Carrasquilla}
\affiliation{Vector Institute for Artificial Intelligence, MaRS Centre, Toronto, ON, Canada M5G 1M1}

\author{Giacomo Torlai}
\affiliation{ Department of Physics and Astronomy, University of Waterloo, Ontario, N2L 3G1, Canada}
\affiliation{Perimeter Institute for Theoretical Physics, Waterloo, Ontario N2L 2Y5, Canada}
\affiliation{Center for Computational Quantum Physics, Flatiron Institute, 162 5th Avenue, New York, NY 10010, USA}

\author{Roger G. Melko}
\affiliation{ Department of Physics and Astronomy, University of Waterloo, Ontario, N2L 3G1, Canada}
\affiliation{Perimeter Institute for Theoretical Physics, Waterloo, Ontario N2L 2Y5, Canada}
\author{Leandro Aolita}
\affiliation{ Instituto de F\'isica, Universidade Federal do Rio de Janeiro, Caixa Postal 68528,
Rio de Janeiro, RJ 21941-972, Brazil}
\affiliation{ ICTP South American Institute for Fundamental Research,
Instituto de F\'isica Te\'orica, UNESP-Universidade Estadual Paulista R.
Dr. Bento T. Ferraz 271, Bl. II, S\~ao Paulo 01140-070, SP, Brazil}

\begin{abstract}

A major bottleneck towards scalable many-body quantum technologies is the
difficulty in benchmarking state preparations, which suffer from an
exponential ``curse of dimensionality'' inherent to the description of
their quantum states.
We present an experimentally friendly method for density matrix
reconstruction based on deep neural-network generative models. The learning procedure
comes with a built-in approximate certificate of the reconstruction and
makes no assumptions about the purity of the state under scrutiny.
It can efficiently handle a broad class of complex systems
including prototypical states in quantum information, as well as ground
states of local spin models common to condensed matter physics.
The key insight is to reduce state tomography to an unsupervised
learning problem of the statistics of an informationally complete 
quantum measurement. This constitutes a modern machine learning
approach to the validation of complex quantum devices, which may in addition prove
relevant as a neural-network Ansatz over mixed states suitable for variational optimization.

\end{abstract}
\maketitle

\baselineskip24pt

\section*{Introduction}
Since the turn of the century, advances in several competing quantum
technologies have demonstrated control and measurements sufficiently accurate
to enable devices  of up to tens, or soon hundreds, of qubits \cite{cirac2012, guzik2012, bloch2012, blatt2012,houck2012, Gross17}.
A comprehensive characterization of modern quantum devices entails the reconstruction
of their quantum state from measurements on identically prepared copies, a
task known as quantum state tomography \cite{VR89,PhysRevA.64.052312, roos04b}.
In their familiar conception, exact tomographic techniques become impractical \cite{haeffner05}
on large quantum systems due to the exponential complexity associated with the description of
generic quantum many-body systems. Physical systems of interest -- such as those
generated by the dynamics of a local Hamiltonian -- are not generic, since their particular
structure guarantees that the full complexity of Hilbert space is in principle not required for their
accurate description \cite{PQSV11, KBGKE11}. Thus, in a wide range of physical situations,
a priori structural information about the state under scrutiny can help alleviate the exponential
scaling. In such cases, a reconstruction can be achieved
by introducing a plausible parametrization of the state, whose computational manipulation and
storage, as well as the number of measurements required for an accurate reconstruction, scale favourably with system size.

Examples following this spirit include permutationally invariant
tomography \cite{Toth10, Moroder12}, compressed sensing \cite{Gross10},
and tomographic schemes based on tensor
networks \cite{cramer2009efficient, MPOtomo, Han2017, Lanyon2017}. Alternatively,
if the target state is known, one can try to certify the fidelity
between the experimental and ideal states, without attempting a reconstruction
of the former \cite{Flammia11,daSilva11,cramer2009efficient, Aolita15, Gluza18}.
All these methods are effective for
different classes of states, but they all share the drawback of limited versatility.
Notably, even though matrix-product-state (MPS) tomography \cite{cramer2009efficient,Han2017, Lanyon2017}
has led to impressive progress in the theoretical and experimental reconstruction
of states of spin chains, generalizations to higher-dimensional lattices
rapidly become computationally intractable. Furthermore, even for one-dimensional
lattices, the entanglement limitations of MPS tomography restricts its application
to states arising from only short-time dynamics \cite{Lanyon2017}.

In the quest for efficiency and versatility, methodologies inspired by undirected
graphical models such as the restricted Boltzmann machine (RBM)~\cite{Torlai2018,Torlai2018B},
as well as based on variational autoencoders\cite{Rocchetto2018} have recently appeared. 
Notably, due to their intrinsically nonlocal structure, RBMs can represent highly entangled 
many-body states using a small number of 
parameters \cite{PhysRevX.7.021021, 2017NatCoGAO,PhysRevB.97.085104, PhysRevX.8.011006}.
However, a scalable formulation for density matrix reconstruction remains elusive. In
current approaches to generalize quantum state tomography to mixed states
through RBM purifications \cite{Torlai2018B}, training and manipulation introduces
exponential scaling in any spatial dimension.

In this paper we combine elements of two state-of-the-art classes of algorithms
to introduce a parametrization of the quantum state that alleviates these scaling
issues. The first is the tensor-network paradigm, designed using well-understood
underlying principles of quantum entanglement to efficiently represent quantum states.
The second and most important is generative models, a key ingredient in modern deep
learning research. These models are used to understand probabilistic distributions
defined over high-dimensional data with rich structure, in tasks such as density estimation,
denoising, missing value imputation, and sampling \cite{Goodfellow-et-al-2016}.
Generative models can be tractably defined and trained in any spatial dimension,
display an extraordinary expressive power \cite{pmlr-v70-raghu17a}, and can represent highly
entangled states \cite{2018arXiv180309780L,PhysRevX.7.021021,2017NatCoGAO,PhysRevB.97.085104}.
We show how to reduce state tomography to an explicit, unsupervised
learning problem using probabilistic models. The reduction consists
of directly parametrizing the outcome probabilities of a tomographically
complete measurement on an arbitrary state with such models. The method is
experimentally friendly since it only requires routinely available
single-particle measurements. We show that this strategy can efficiently
learn a variety of complex states, from paradigmatic multi-qubit states
undergoing local noise, to ground states of local spin models in both one
and two spatial dimensions. Our approach is also reliable since the state reconstruction
can be approximately certified efficiently by sampling from the reconstructed distribution.

\section*{Neural representation of general quantum states and the learning strategy}

\begin{figure}
\centering
\includegraphics[width=5.0in]{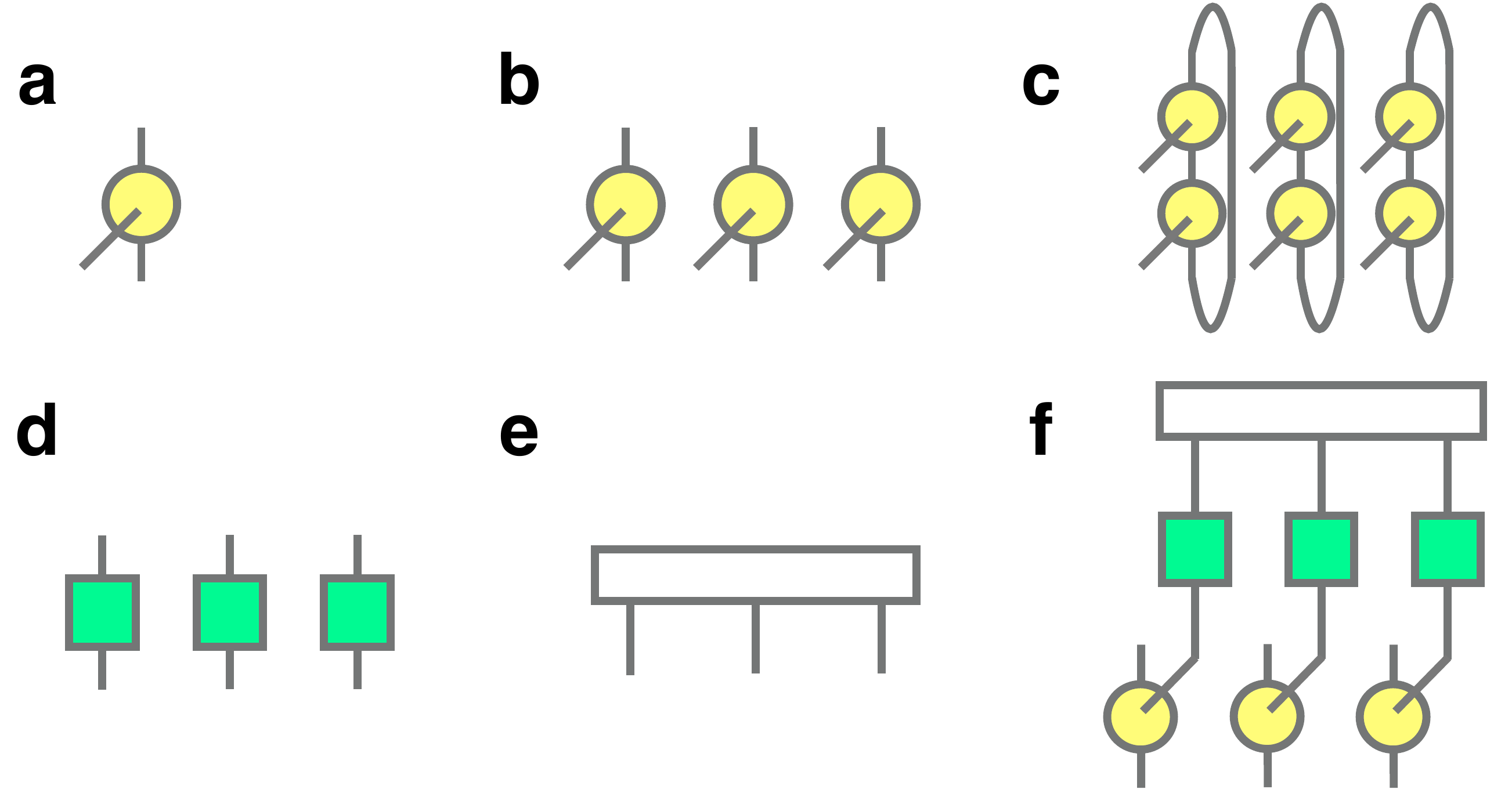}\newline
\caption{Tensor-network schematics of the formalism.
(a) The single-qubit measurement $\boldsymbol{M}=\{M^{(a)}\}_a$
is a three-index tensor represented by a yellow circle with three
 emerging lines. Vertical indices act on the physical degrees
of freedom while the outcoming one labels the measurement outcome $a$.
(b) The  $N$-qubit measurement
$\boldsymbol{M}=\big\{M^{(a_1)}\otimes M^{(a_2)}\otimes \hdots M^{(a_N)}\big\}_{a_1, \hdots a_N}$
corresponds to the same local measurement on each qubit.
(c) The components $T_{\boldsymbol{a},\boldsymbol{a}'} =
\text{Tr}\left[  M^{(\boldsymbol{a})} M^{(\boldsymbol{a}')} \right]$ of the
overlap matrix $\boldsymbol{T}$ correspond to the physical-index contraction
of the measurement operators. (d) Consequently, when $\boldsymbol{T}$
is invertible, its inverse $\boldsymbol{T}^{-1}$ factorizes into the
tensor product of single-qubit matrices $\boldsymbol{T}_{\text{1-qubit}}^{-1}$
(green squares). (e) We represent the outcome distribution $\boldsymbol{P}$
as an $N$-index tensor where the indices encode the measurement outcome of
each qubit. (f) A generic density matrix $\varrho$ is given by a contraction
of $\boldsymbol{P}$, $\boldsymbol{T}^{-1}$, and $\boldsymbol{M}$ over the
outcome indices: $\varrho = \sum_{\boldsymbol{a},\boldsymbol{a}'}
P(\boldsymbol{a}) \,T^{-1}_{\boldsymbol{a},\boldsymbol{a}'}\,M^{(\boldsymbol{a}')}=
\mathbb{E}_{\boldsymbol{a}\sim\boldsymbol{P}}\left(  \sum_{\boldsymbol{a}'} T^{-1}_{\boldsymbol{a},\boldsymbol{a}'}\,M^{(\boldsymbol{a}')} \right)$.
Expressed this way, all non-local correlations of $\varrho$ are encoded
explicitly into $\boldsymbol{P}$, as both $\boldsymbol{T}$ and $\boldsymbol{M}$
are single-qubit factorable. Our state Ansatz parameterizes the outcome
distribution  with a neural-network generative model, i.e. $\boldsymbol{P}= \boldsymbol{P}_{\text{model}}$.
\label{fig:TNrep}}
\end{figure}

We consider measurements given by informationally complete (IC)
positive-operator valued measures (POVMs) (see Methods). POVMs describe the
most general type of measurements allowed by quantum theory, beyond the usual
notion of von Neumann projective measurements~\cite{Nielsen}.
Informational completeness means that the measurement statistics
contains all of the information about the state, thus specifying it univocally.
We consider physical systems composed of $N$ qubits and build
our measurements starting from an $m$-outcome single-qubit POVM
$\boldsymbol{M}=\{M^{(a)}\}_a$ [see Fig.\ref{fig:TNrep} (a)], defined by positive
semi-definite operators $M^{(a)}\geq 0$, each one labeled by a measurement outcome
$a=0,1,..,m-1$. These satisfy the normalization condition $\sum_a M^{(a)}=\openone$.
Our $N$-qubit measurement is given by the tensor product of the single-qubit POVM
elements $\boldsymbol{M}=\big\{M^{(a_1)}\otimes M^{(a_2)}\otimes \hdots M^{(a_N)}\big\}_{a_1, \hdots a_N}$,
graphically depicted in Fig~\ref{fig:TNrep} (b).

By virtue of Born's rule, the probability distribution $\boldsymbol{P}=\{P(\boldsymbol{a})\}_{\boldsymbol{a}}$ over measurement outcomes $\boldsymbol{a}=(a_1,a_2,...,a_N)$  on a quantum state $\varrho$, with
$P(\boldsymbol{a})\geq 0$ and $\sum_{\boldsymbol{a}}P(\boldsymbol{a})=1$, is given by the linear expression
$P(\boldsymbol{a})=\text{Tr}\left[ M^{(\boldsymbol{a})}\, \varrho\right]$.
Provided that the measurement is informationally complete, this relation can formally
be inverted. In other words, the density matrix can be unambiguously inferred
from the probability distribution of measurement outcomes. This can be explicitly
expressed in a concise way when the overlap matrix
$\boldsymbol{T}$, of elements $T_{\boldsymbol{a},\boldsymbol{a}'} =
\text{Tr}\left[  M^{(\boldsymbol{a})} M^{(\boldsymbol{a}')} \right]$, is invertible:
$\varrho = \sum_{\boldsymbol{a},\boldsymbol{a}'} P(\boldsymbol{a})\,
T^{-1}_{\boldsymbol{a},\boldsymbol{a}'}\,M^{(\boldsymbol{a}')}=\mathbb{E}_{\boldsymbol{a}\sim\boldsymbol{P}}\left( \sum_{\boldsymbol{a}'} T^{-1}_{\boldsymbol{a},\boldsymbol{a}'}\,M^{(\boldsymbol{a}')} \right)$, with $\mathbb{E}_{\boldsymbol{a}\sim\boldsymbol{P}}$
representing the expectation value over $\boldsymbol{a}$ distributed according to
$\boldsymbol{P}$ [see Figs.~\ref{fig:TNrep} (c-f) for a graphical representation of these relations].
Notice that in Fig.~\ref{fig:TNrep} we use the language of tensor networks and its
graphical notation, first introduced by Penrose~\cite{penrose1971}, to pictorially
reason about the elements and structure of the Ansatz.

Given a collection of experimental outcomes
$E=\{\boldsymbol{a}_1, \boldsymbol{a}_2,\dotsc,\boldsymbol{a}_{N_{s}}\}$, with $N_s$ samples,
our strategy begins with learning a model $P_{\text{model}}(\boldsymbol{a})$
that describes the measurement statistics in terms of expressive neural generative
models. This task can be carried out using a wide variety of models and training
strategies, including variational autoencoders, generative adversarial networks, restricted
Boltzmann machines, and powerful autoregressive models based upon recurrent
neural networks (RNN), among others. To demonstrate our approach, we first consider the prototypical
$N$-qubit Greenberger-Horne-Zeilinger (GHZ) state, which is a highly non-classical state specified by
$|\Psi\rangle= \frac{1}{\sqrt{2}}\left(|0\rangle^{\otimes N} +|1\rangle ^{\otimes N}  \right)$.
We examine mixed states arising from GHZ states subject to local depolarizing
noise on each qubit independently.
Each qubit is depolarized (i.e., all its information lost)
with probability $p$, while it is untouched with probability $1 - p$~\cite{Nielsen}.
We measure the so-called tetrahedral POVM $\boldsymbol{M}_{\text{tetra}}$ on each qubit.
This is an IC POVM with $m=4$ outcomes, which is the minimum required for informational
completeness, and an invertible overlap matrix. Each measurement operator in
$\boldsymbol{M}_{\text{tetra}}$ is proportional to a rank-1 projector pointing to a
different vertex of a regular tetrahedron in the Bloch sphere (see Methods).

\section*{Numerical experiments}

\begin{figure}
\centering
\includegraphics[width=6.6in]{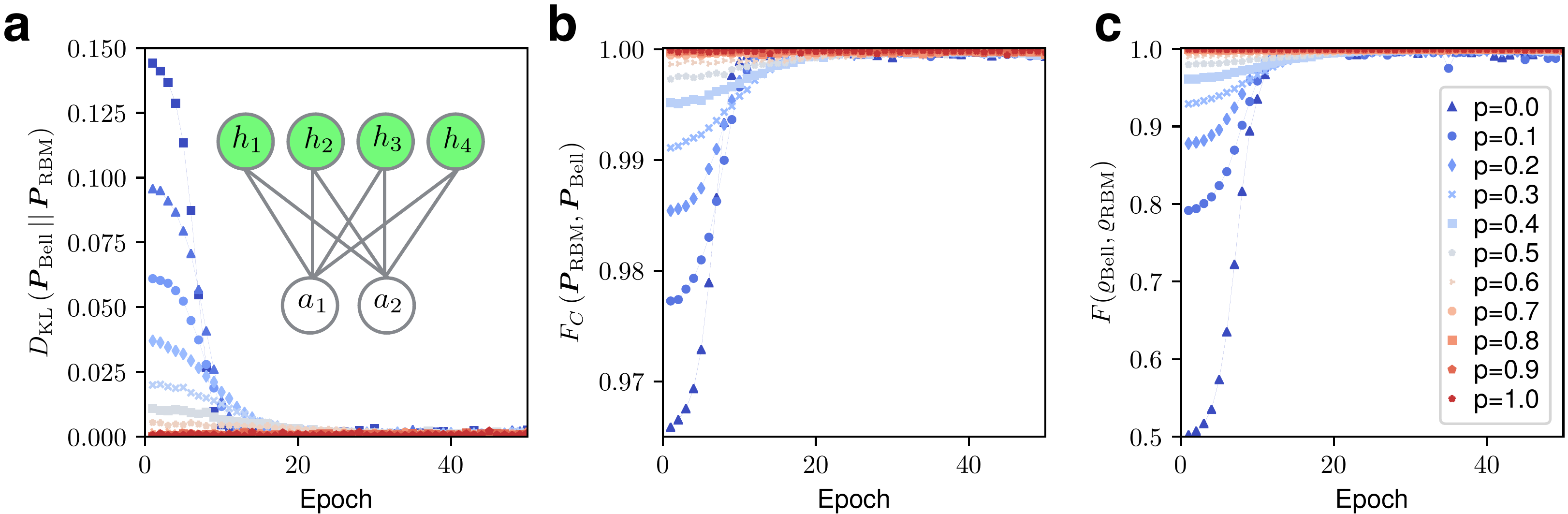}
\caption{
Learning a Bell state under local depolarizing noise. (a) The Kullback-Liebler divergence
$D_{\text{KL}}$ as a function
of the training epochs for different depolarization probabilities $p$. One epoch refers to a training cycle
that exposes the learning algorithm to the entire training dataset. The sample size of the dataset used
during the training is $N_s=6\times10^{4}$ observations. The inset displays a graphical representation
of the RBM employed in the reconstruction of the state. This architecture features a visible layer with
multinomial variables $a_i=0,1,2,3$ (white circles) and binary hidden units $h_i=0,1$ (green) connected
via a set of weight parameters represented by the grey connections.(b) The classical fidelity $F_C$
between the reconstructed distribution
$\boldsymbol{P}_{\text{RBM}}$ and $\boldsymbol{P}_{\text{GHZ}}$. (c) The quantum fidelity $F$ between
the depolarized Bell state and its RBM reconstruction.\label{fig:2qubit} }
\end{figure}

To begin the implementation of our method, we require a parameterization of $P_{\text{model}}(\boldsymbol{a})$.
Because of their extensive familiarity as a tool to represent quantum states
\cite{Carleo602,PhysRevB.97.085104,Torlai2018,Torlai2018B,
PhysRevX.7.021021,2017NatCoGAO,PhysRevB.96.195145,PhysRevB.96.205152,1751-8121-51-13-135301},
we parametrize $P_{\text{model}}(\boldsymbol{a})$ in terms of an RBM as a first demonstration.
The main extension over traditional RBMs is the need for
four-dimensional multinomial visible units, which were previously introduced in the context of collaborative filtering~\cite{Salakhutdinov2007},
for applications such as the famous Netflix Prize~\cite{Bennett07thenetflix}.
Using such RBMs, we learn the statistics of numerically simulated measurements on the GHZ
state with $N=2$ qubits, i.e.~the Bell state $|\Psi\rangle= \frac{1}{\sqrt{2}}\left(|0\rangle\otimes|0\rangle +|1\rangle\otimes|1\rangle  \right)$, under
local depolarization with different values of noise strengths $0\leq p\leq 1$. Our results are shown in Fig.~\ref{fig:2qubit}.

The RBM model is trained using a standard contrastive divergence
procedure~\cite{Goodfellow-et-al-2016}, which aims at maximizing the
log-likelihood of the the data with respect to the parameters in the model.
In Fig.~\ref{fig:2qubit} (a) we show the Kullback-Leibler (KL)
divergence $D_{\text{KL}}\left(\boldsymbol{P}_{\text{Bell}} \left|\right| \boldsymbol{P}_{\text{RBM}} \right)=
\mathbb{E}_{\boldsymbol{a}\sim\boldsymbol{P}_{\text{Bell}}}\left[ \log{\frac{P_{\text{RBM}}
(\boldsymbol{a})}{P_{\text{Bell}}(\boldsymbol{a})}} \right]$, which measures
how much the model distribution $\boldsymbol{P}_{\text{RBM}}$ diverges
from the exact one $\boldsymbol{P}_{\text{Bell}}$.
As the training progresses, the KL divergence successfully decreases
to values near zero for all values of noise $p$. Likewise, the classical
fidelity $F_{C}\left( \boldsymbol{P}_{\text{RBM}},\boldsymbol{P}_{\text{Bell}}\right) =
\mathbb{E}_{\boldsymbol{a}\sim\boldsymbol{P}_{\text{Bell}}}\left[
\sqrt{\frac{P_{\text{RBM}}(\boldsymbol{a})}{P_{\text{Bell}}(\boldsymbol{a})}}  \right]
= \mathbb{E}_{\boldsymbol{a}\sim\boldsymbol{P}_{\text{RBM}}}\left[
\sqrt{\frac{P_{\text{Bell}}(\boldsymbol{a})}{P_{\text{RBM}}(\boldsymbol{a})}}  \right]$
is a standard measure of proximity between two distributions. Indeed, it is such that
$0\leq F_{C}\left( \boldsymbol{P}_{\text{model}},\boldsymbol{P}_{\text{Bell}} \right) \leq 1$
for all $\boldsymbol{P}_{\text{RBM}}$ and $\boldsymbol{P}_{\text{Bell}}$, with
$F_{C}\left( \boldsymbol{P}_{\text{RBM}},\boldsymbol{P}_{\text{Bell}}  \right) = 1$
if and only if $\boldsymbol{P}_{\text{RBM}}=\boldsymbol{P}_{\text{Bell}}$.
As depicted in Fig.~\ref{fig:2qubit}(b), this quantity approaches unity as  $\boldsymbol{P}_{\text{RBM}}$ converges toward $\boldsymbol{P}_{\text{Bell}}$.  Finally, in Fig.~\ref{fig:2qubit}(c)
we consider the quantum fidelity $F(\varrho_{\text{Bell}},\varrho_{\text{RBM}})=
\text{Tr}\left[\sqrt{\sqrt{\varrho_{\text{Bell}}}\,\varrho_{\text{RBM}}\,\sqrt{\varrho_{\text{Bell}}}}\right]$,
the standard measure of proximity between states that generalizes $F_C$ from
probability distributions to density matrices.

In general, the bound
$F_{C}\left( \boldsymbol{P}_{\text{RBM}}, \boldsymbol{P}_{\text{Bell}} \right) \geq F(\varrho_{\text{Bell}},\varrho_{\text{RBM}})$
holds. However, in our numerical experiments we observe that both the classical
and quantum fidelities collectively approach one with remarkably similar
behavior as a function of the training epoch. This suggests that the
classical fidelity serves as good figure of merit for the quality of
the reconstructed quantum state.
We argue that informational completeness together with the physical nature of
the state should -- for most practical situations -- imply that if $F_{C}$ is close to one, then so is $F$.
From an experimental perspective it is particularly natural to
consider the fidelity between $\boldsymbol{P}$ and $\boldsymbol{P}_{\text{model}}$
as an indication for reconstruction quality, given that the experimental observables
correspond to samples drawn from $\boldsymbol{P}$. Whereas $F$ is in practice inaccessible
for anything but modest $N$, $F_{C}$ can be efficiently estimated accurately for
large $N$ via Monte Carlo, as an average over samples from
$\boldsymbol{P}_{\text{model}}$, if $\boldsymbol{P}$ is known.

We now focus on the scaling of the resources required for the learning procedure
as a function of the number of qubits $N$. First we observe that, for qubit numbers
$N>4$, the training becomes increasingly difficult for the RBM in the regime of
small noise $p$. We therefore opt for a different probabilistic model based upon autoregressive recurrent
neural networks (see Appendix). These models are ordinarily used in end-to-end
sequence learning tasks~\cite{NIPS2014_5346}, and are the state-of-the-art engines behind
machine translation and speech recognition systems~\cite{45610,46687}.
We discovered that, in our setting, RNN models are faster to train compared to RBMs.

\begin{figure}
\centering
\includegraphics[width=5.4in]{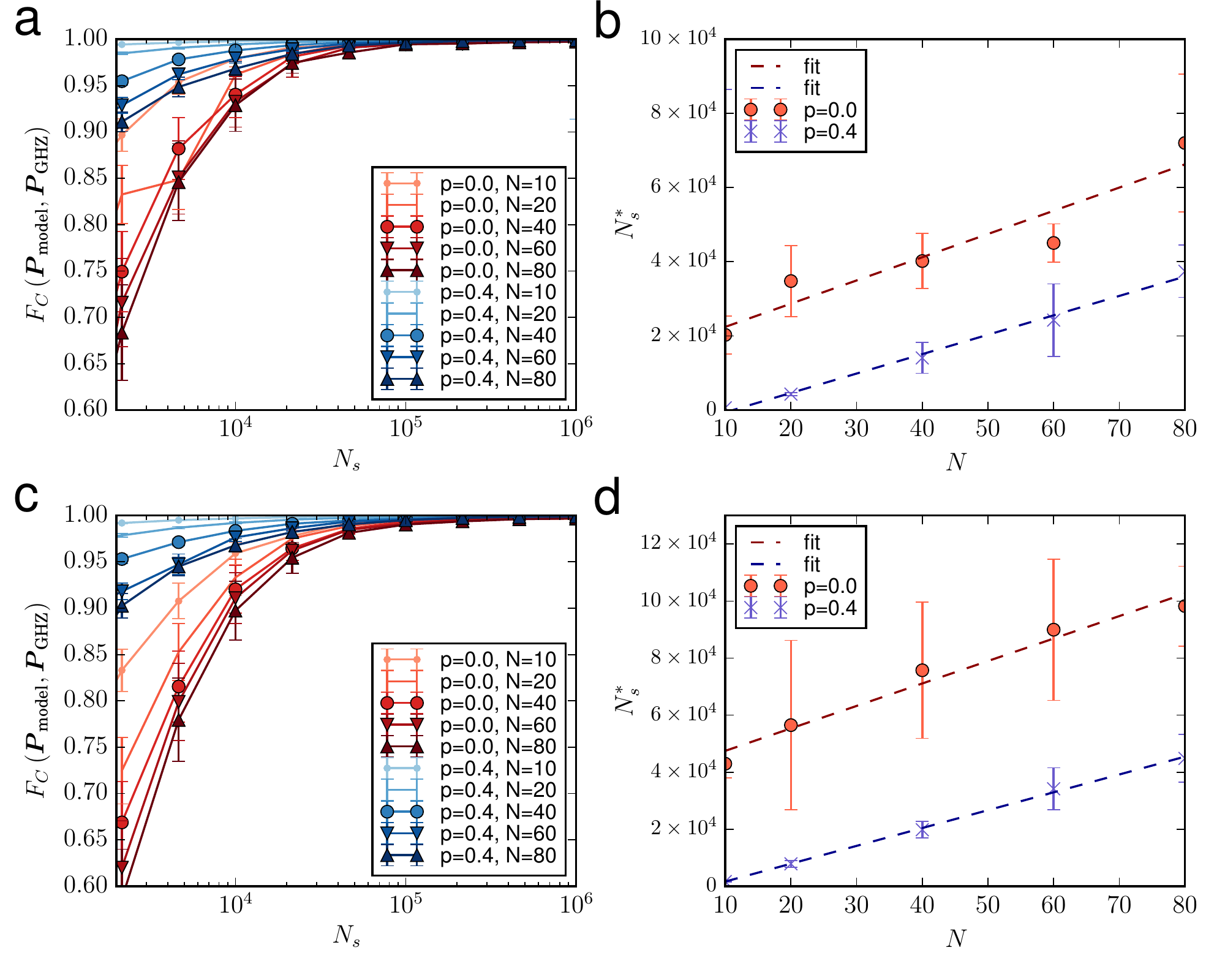}
\caption{
Sample complexity of learning locally depolarized GHZ states with recurrent neural-network (RNN) models.
(a) Estimated average classical fidelity as a function of the number of measurements performed on the
GHZ state for two different values of noise $p=0$ (red) and $p=0.4$ (blue) using the
4-outcome tetrahedral POVM. The reported classical fidelity is an average over a few ($N_m=30$)
models taken around the optimal model during training and the error bars represent
the one s.d. statistical uncertainty calculated over the different models. The classical
fidelity for each model during training is estimated based upon $10^5$ samples drawn from the RNN model.
(b) The number of samples $N_s^*$ necessary to attain an average classical fidelity
$F_C\left(\boldsymbol{P}_{\text{RNN}},\boldsymbol{P}_{\text{GHZ}}\right)=0.99$. (c) and (d)
are analogous to (a) and (b) for the 6-outcome Pauli POVM. Dashed lines represent linear fits,
all of which exhibit a correlation coefficient $r>0.94$.
\label{fig:samplecomplex} }
\end{figure}

To study noisy GHZ states for large $N$, we produce synthetic datasets mimicking
experimental measurements of the exact distribution $\boldsymbol{P}_{\text{GHZ}}$ (see Appendix for details).
For the tetrahedral POVM, we investigate the classical fidelity as a function of $N_s$ used for
training a RNN model $\boldsymbol{P}_{\text{RNN}}$ for different values of $N$ and $p$.
We plot the results in Fig.~\ref{fig:samplecomplex}(a). As in the RBM case,
we find that $F_C\left(\boldsymbol{P}_{\text{RNN}},\boldsymbol{P}_{\text{GHZ}}\right)$ quickly approaches
unity for all of the states that we consider. We also find that learning the noiseless ($p=0$) states takes
significantly more effort in terms of training set size. This reflects
the larger amount of information contained in a pure state relative to
its depolarized counterparts \cite{Nielsen,PreskillNotes}. Remarkably, as shown in
Fig.~\ref{fig:samplecomplex}(b), the number $N_s^*$ of samples required to learn the state up to
$F_C\left(\boldsymbol{P}_{\text{RNN}},\boldsymbol{P}_{\text{GHZ}}\right)=0.99$ is found to
scale approximately linearly with $N$.
To investigate if this scaling is a just a peculiarity of the particular POVM chosen,
we implement the same learning protocol with a different measurement. We use a single-qubit
IC POVM with $m=6$ outcomes, each one described by a POVM element proportional to the
rank-1 projector onto one of the 6 eigenstates of the three usual Pauli matrices (see Methods).
We refer to this as the Pauli-6 POVM and denote it by $\boldsymbol{M}_{\text{Pauli}-6}$.
Even though the $N$-qubit sample space is now exponentially larger than the tetrahedral
POVM, we find out that $N_s^*$ still scales linearly with $N$, with only a slightly
larger slope than for $\boldsymbol{M}_{\text{tetra}}$. While we again define $N_s^*$ based
on an average value of $F_C\left(\boldsymbol{P}_{\text{RNN}},\boldsymbol{P}_{\text{GHZ}}\right)=0.99$,
we remark that in all cases, the maximum classical fidelity we find is at least
$F_{C}\left( \boldsymbol{P}_{\text{RNN}},\boldsymbol{P}_{\text{GHZ}}\right)\geq 0.999$ for a number of samples
$N_s=10^{6}$ for all system sizes and values of noise $p$.
Finally, we note that this sample complexity
is consistent with arguments that, in a setting where measurements are chosen probabilistically, quantum
states can be approximately learned using only a linear number of measurements \cite{Aaronson3089,2017arXiv171200127R}.

\begin{figure}
\centering
\includegraphics[width=5.4in]{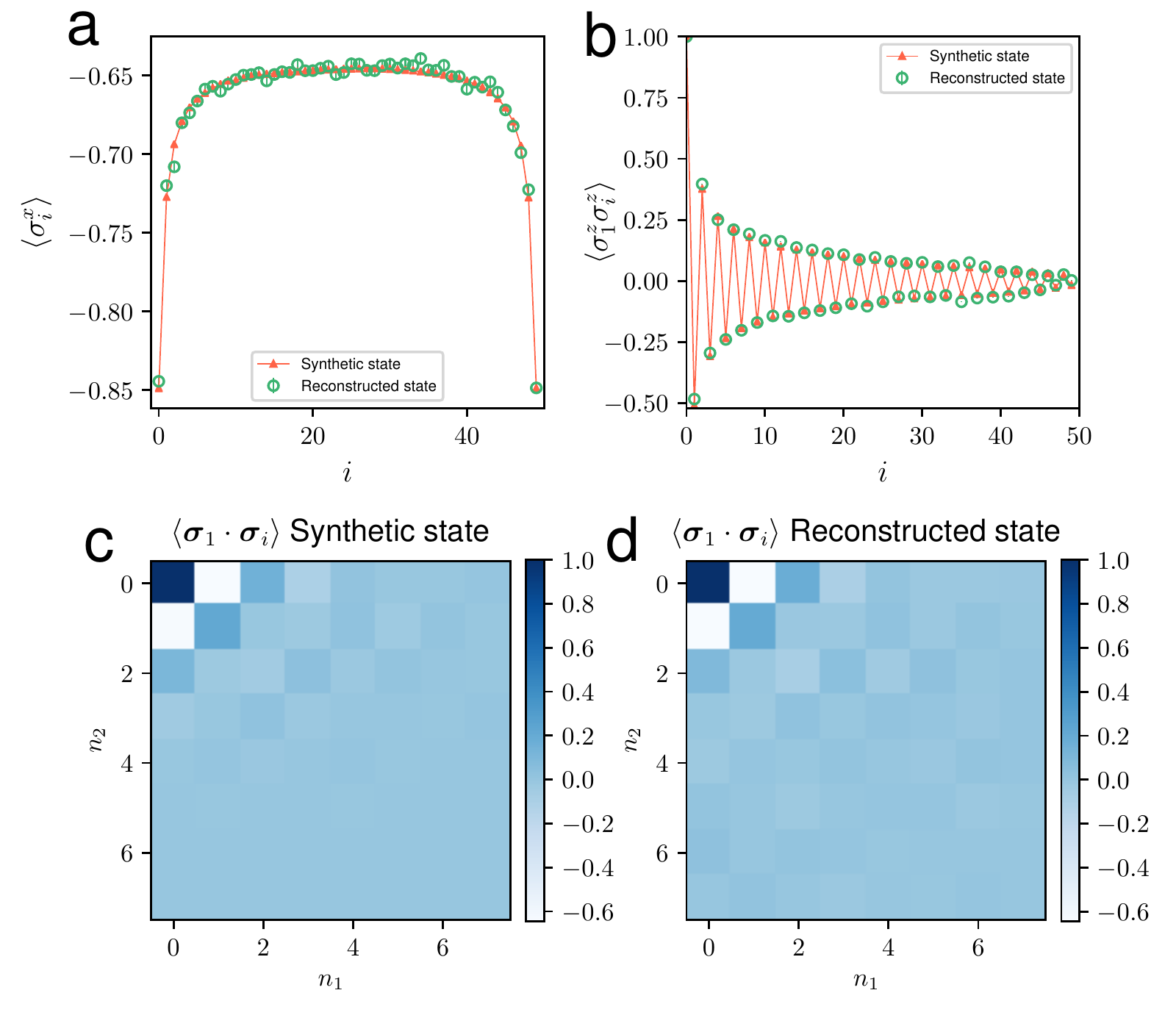}
\caption{
Learning ground states of local Hamiltonians in one and two dimensions with RNN models for the Pauli-4 POVM ($m=4$ outcomes per qubit).
Similar results are obtained for the tetrahedral POVM. Antiferromagnetic transverse-field Ising model in 1D for $N=50$ spins:
(a) 1-body $\langle \sigma^x_i \rangle$  and (b) 2-body $\langle \sigma^{z}_1 \sigma^{z}_i \rangle$ correlators calculated
from the exact synthetic state and from the RNN reconstruction as a function of the lattice distance $i$.
Antiferromagnetic translationally-invariant Heisenberg model on the triangular lattice with $N=8\times 8$ spins:
2-body correlators $\langle \boldsymbol{\sigma}_1 \boldsymbol{\sigma}_i \rangle$
calculated from (c) the exact synthetic state and (d) from the RNN reconstruction as a function of the position of the different lattice
sites $i$. Here, $i = n_1\times L + n2 $ labels the $i$-th spin located at position $\boldsymbol{r}_i = n_1 \boldsymbol{a_1} + n_2 \boldsymbol{a_2}$,
where $\boldsymbol{a}_1=(1,0)$ and $\boldsymbol{a}_2=\frac{1}{2}(1,\sqrt{2})$  are primitive lattice vectors of the triangular lattice,
and $n_{i=1,2}=0,1,\ldots L-1$.}
\label{fig:localH}
\end{figure}

We now turn our attention to the reconstruction of states arising from ground states of
Hamiltonians with local interactions, relevant for modern condensed matter physics, quantum
chemistry, atomic and molecular optics, and quantum computing.
We begin with the paradigmatic example of the antiferromagnetic
transverse-field Ising model in one dimension, with Hamiltonian
$H=J\sum_{\langle ij\rangle }\sigma_{i}^{z}\sigma_{j}^{z}+h\sum_{i}\sigma_{i}^{x}$.
The transverse-field Ising model is the testbed par excellence for state-of-the-art quantum simulators in both strongly
out-of-equilibrium dynamics \cite{Barends15, Lanyon2017, Gross17} as well as quasi-adiabatic
regimes \cite{Friedenauer08, Kim10, Islam11, Simon11, 2017Natur.551..579B}. It has been
demonstrated in ion traps \cite{Friedenauer08, Kim10, Islam11,Lanyon2017}, cold atoms in optical
lattices \cite{Simon11, Gross17}, superconducting qubit circuits \cite{houck2012, Barends15}
of up to 1800 qubits \cite{2018Natur.560..456K}, and Rydberg-atom platforms of up to
51 qubits \cite{2017Natur.551..579B}.
We focus on the most challenging parameter regime for state reconstruction,
the quantum critical point at $h/J=1$. For a system of $N=50$ qubits, we first obtain
synthetic data mimicking $10^6$  experimental POVM measurements on the ground state of $H$,
then perform a reconstruction of the state from these measurements using the same RNN model
used in Fig.~\ref{fig:samplecomplex}. Here, we test our method
with two different IC POVMs, the tetrahedron and a modified Pauli POVM (termed Pauli-4 POVM,
$\boldsymbol{M}_{\text{Pauli}-4}$). The latter can be experimentally implemented
with the same ease as Pauli-6 but with the advantages of having only $m=4$ outcomes and an
invertible overlap matrix (see Methods). Invertibility of $\boldsymbol{T}$ is important
because it gives us the remarkable ability to efficiently estimate expectation values of local observables
on the reconstructed state $\varrho_{\text{RNN}}$ directly from $\boldsymbol{P}_{\text{RNN}}$, i.e.
without an explicit construction of $\varrho_{\text{RNN}}$. This is done stochastically via sampling from
$\boldsymbol{P}_{\text{RNN}}$ (see Methods). The resulting reconstructions attain a fidelity
$F_{C}\left(\boldsymbol{P}_{\text{RNN}}, \boldsymbol{P}_{\text{Ising}}\right)\approx 0.998$.
Furthermore, we find that the 1- and 2-body correlations functions (and, therefore, also the total energy)
of $\varrho_{\text{RNN}}$ display an excellent agreement with the synthetic state,
consistent with the exact values to within error bars (Figs.~\ref{fig:localH} (a) and (b)).

The results we have presented so far demonstrate the power and scalability of our technique
on some states with simple structure,
which are also amenable to previous tomography
approaches \cite{cramer2009efficient,Han2017, Lanyon2017,Torlai2018,Torlai2018B}.
We now expand our investigation to a case that lies outside of conventional approaches.
We study a model of frustrated magnetism in two spatial dimensions whose ground state
possesses a highly non-trivial sign structure in the computational basis: the antiferromagnetic
Heisenberg model in the triangular lattice with Hamiltonian
$H = \sum_{\langle i j \rangle} {\bm \sigma}_i \cdot {\bm \sigma}_j$, where
${\bm \sigma}_i=\left(\sigma^{x}_i, \sigma^{y}_i,\sigma^{z}_i \right)$ is the Pauli vector at site $i$.
We use a tensor network approximation to the ground state of the model for a lattice of $N=L \times L = 8\times8=64$
spins to produce synthetic samples from the tetrahedral and Pauli-4 POVM measurements (see Appendix for details).
We find that, surprisingly, the RNN model learns the state with a similar success
as in the one-dimensional case. Apart from the classical fidelity, which, e.g., for the
Pauli-4 measurement, reaches
$F_{C}\left(\boldsymbol{P}_{\text{RNN}},\boldsymbol{P}_{\text{Heisenberg}}\right)\approx 0.98$, we
observe a remarkable agreement between the correlation function
$\langle \boldsymbol{\sigma}_1 \cdot \boldsymbol{\sigma}_i \rangle$ measured
from our reconstruction $\varrho_{\text{RNN}}$ and the exact value in the synthetic state (see Figs.~\ref{fig:localH} (c) and (d)).

We take advantage of this example to emphasize that our state ansatz corresponds to a
contraction between a neural function approximator parametrizing $\boldsymbol{P}$ and
a Kronecker factorized tensor network composed of local, complex-valued
tensors, which are efficiently contractible. The sign structure of the state
is provided by the factorized tensor, weighted by the distribution $P(\boldsymbol{a})$.
All of the entanglement, and  -- more generally -- any potential classical intractability
in the state, can be directly traced back to $\boldsymbol{P}$.
One can readily apply any other  powerful state-of-the-art probabilistic model and training
strategy, such as, e.g., the variational autoencoder, generative adversarial networks, and
generic factor graphs, all of which have natural definitions in higher spatial dimensions. These and
other models may lead to representations of quantum states beyond those considered in this manuscript.
Compared to previous approaches~\cite{Carleo602,Torlai2018,Torlai2018B}, which rely on
complex-valued generalizations of graphical models, our neural networks contain only
real parameters, which makes them faster to train. Thus, our ansatz may prove applicable
in the study of ground and thermal states of quantum many-body systems via variational
energy minimization.

\section*{Conclusions and outlook}
In conclusion, these results demonstrate a scalable machine learning procedure for reconstructing
pure and mixed states with structure described by a wide range of generative models
in conjunction with easily available measurements. The procedure includes a certification
scheme based upon the classical fidelity of the measurement statistics.
We demonstrated our method for prototypical states in quantum information as well as ground
states of local Hamiltonians relevant to condensed matter, cold atomic systems,
and quantum simulators. Most importantly, this work demonstrates how state-of-the-art
algorithms in machine learning, combined with established tensor network frameworks
in quantum physics, may help make headway in validating and characterizing
quantum simulators and commercially available quantum devices in the near-term era of approximate
quantum computing \cite{2018arXiv180100862P}.

\section*{Methods}

\section{Informationally complete generalized measurements}
\label{sec:POVM}
Generalized (non von Neumann) measurements are described by positive-operator valued measures (POVMs).
These are defined by decompositions $\boldsymbol{M}=\{M^{(a)}\}_a$ of the identity $\openone$, i.e.
$\sum_a M^{(a)}=\openone$, in terms of non-negative operators $M^{(a)}\geq 0$. Additionally, a
POVM $\boldsymbol{M}$ is said to be informationally complete (IC) if its elements $M^{(a)}$ span
the whole space of bounded-norm, linear operators on the Hilbert in question, where the density
matrix to reconstructs lives. Here we consider 3 different single-qubit IC POVMs.

The first one is the tetrahedral POVM
$\boldsymbol{M}_{\text{tetra}}=\big\{M^{(a)}=\frac{1}{4}(\openone+\boldsymbol{s}^{(a)}\cdot\boldsymbol{\sigma})\big\}_{a\in\{0,1,2,3\}}$,
whose outcomes correspond to sub-normalized rank-1 projectors along the directions
$\boldsymbol{s}^{(0)}=(0,0,1)$, $\boldsymbol{s}^{(1)}=(\frac{2\sqrt{2}}{3},0,-\frac{1}{3})$,
$\boldsymbol{s}^{(2)}=(-\frac{\sqrt{2}}{3},\sqrt{\frac{2}{3}},-\frac{1}{3})$,
and $\boldsymbol{s}^{(3)}=(-\frac{\sqrt{2}}{3},-\sqrt{\frac{2}{3}},-\frac{1}{3})$ in the Bloch sphere.
These define a regular tetrahedron, which explains the name and also renders
$\boldsymbol{M}_{\text{tetra}}$ symmetric. A POVM is symmetric if the overlap between any
two different elements is the same, i.e. if its overlap matrix has constant non-diagonal elements.
The overlap matrix
\begin{equation}
\boldsymbol{T}_{\text{tetra}}=
\left[ {\begin{array}{cccc}
   \frac{1}{4} & \frac{1}{12} & \frac{1}{12} & \frac{1}{12} \\
   \frac{1}{12} & \frac{1}{4} & \frac{1}{12} & \frac{1}{12} \\
   \frac{1}{12} & \frac{1}{12} & \frac{1}{4} & \frac{1}{12} \\
    \frac{1}{12} & \frac{1}{12}  & \frac{1}{12} & \frac{1}{4} \\
    \end{array} } \right]
\end{equation}
of $\boldsymbol{M}_{\text{tetra}}$ is invertible. The experimental implementation
of $\boldsymbol{M}_{\text{tetra}}$ relies on Neumark's dilation theorem \cite{peres95}.
This states that any POVM composed of $m$ rank-1 operators on a $d$-dimensional Hilbert
space, for $d\leq m$, can be accomplished by a properly crafted projective measurement
in an extended Hilbert space with dimension $m$. This implies that $\boldsymbol{M}_{\text{tetra}}$
can be physically realized by coupling the system qubit to an ancillary qubit and performing
a von Neumann measurement on the two qubits (see, e.g., Ref. \cite{PhysRevA.86.062107} for explicit constructions).

The second example is the Pauli-6 POVM $\boldsymbol{M}_{\text{Pauli}-6}$ with $m=6$
outcomes. Each one of its elements is, again, a sub-normalized rank-1 projector:
$\boldsymbol{M}_{\text{Pauli}-6} =\big\{M^{(0)} = \frac{1}{3}\times\ketbra{0}{0},
M^{(1)}= \frac{1}{3}\times\ketbra{1}{1}, M^{(2)}= \frac{1}{3}\times\ketbra{+}{+}, M^{(3)}= \frac{1}{3}\times\ketbra{-}{-}, M^{(4)}= \frac{1}{3}\times\ketbra{r}{r},
M^{(5)}= \frac{1}{3}\times\ketbra{l}{l}\big\}$, where $\{\ket{0}, \ket{1}\}$, $\{\ket{+}, \ket{-}\}$, and $\{\ket{r},\ket{l}\}$
stand for the eigenbases of the Pauli operators $\sigma^{z}$, $\sigma^{x}$, and $\sigma^y$, respectively.
Hence, this POVM encapsulates into a single (generalized) measurement all three usual
(von Neumann) Pauli measurements. Therefore, it can be physically implemented directly
(with no ancilas) by first randomly choosing $x$, $y$, or $z$, and then measuring the
respective Pauli operator. Clearly, any positive probabilities other than $\frac{1}{3}$ are
possible too. This POVM is not symmetric, and its overlap matrix is
\begin{equation}
\boldsymbol{T}_{\text{Pauli}-6}=
\frac{1}{9}\left[ {\begin{array}{cccccc}
   1 & 0 & 1/2& 1/2 & 1/2 & 1/2\\
   0 & 1 & 1/2& 1/2 & 1/2 & 1/2\\
   1/2 & 1/2 & 1 & 0 & 1/2 & 1/2\\
    1/2& 1/2  & 0 & 1  & 1/2 & 1/2\\
    1/2 & 1/2 & 1/2 & 1/2 & 1 & 0\\
    1/2 & 1/2 & 1/2 & 1/2 & 0 & 1\\
    \end{array} } \right],
\end{equation}
which is not invertible. This means that the linear inversion of Figs.~\ref{fig:TNrep} (d)
and (f) is not possible. It also implies that estimations of expectation values of local
observables are difficult even for moderate $N$. Luckily, however, non-invertibility
poses no particular challenge for the estimation of the classical fidelity, as discussed in the main text.

The third single-qubit POVM we consider is the Pauli-4 POVM
$\boldsymbol{M}_{\text{Pauli}-4}=\big\{M^{(0)} = \frac{1}{3}\times\ketbra{0}{0},
M^{(1)}= \frac{1}{3}\times\ketbra{+}{+}, M^{(2)}= \frac{1}{3}\times\ketbra{r}{r},
M^{(3)}= \openone - M^{(0)}- M^{(1)}-  M^{(2)} \big\}$.
The Pauli-4 can be physically implemented with the same ease as Pauli-6. In
fact, the experimental procedure is almost the same, the only difference being that,
for Pauli-4, one adds a trivial classical post-processing that identifies three different
outcomes ($a=1$, $a=3$, and $a=5$) of Pauli-6 with a single one ($a=3$) of Pauli-4. Pauli-4 
is neither symmetric nor rank-1. However, apart from the computational advantage of having only $m=4$ outcomes, its overlap matrix
\begin{equation}
\boldsymbol{T}_{\text{Pauli}-4}=
\frac{1}{9}\left[ {\begin{array}{cccc}
   1   & 1/2 & 1/2 & 1 \\
   1/2 & 1   & 1/2 & 1 \\
   1/2 & 1/2 & 1   & 1 \\
   1   & 1   & 1 &   6 \\
    \end{array} } \right].
\end{equation}
is invertible.
\section{Direct stochastic estimation of local observables}
Given samples $N_s$ from a model distribution $\boldsymbol{P}_{\text{model}}$, parametrizing a
state $\varrho_{\text{model}}$, and a single-qubit factorable POVM with invertible
overlap matrix, it is possible to efficiently compute the expectation value
$\langle O \rangle=\text{Tr}\left[ O \, \varrho_{\text{model}}\right]$ of any
observables $O$ acting non-trivially only on a constant (i.e. $N$-independent) number
of qubits. Remarkably, this can be done directly from the samples, i.e. without the
reconstructed density matrix $\varrho_{\text{model}}$. Particular examples of such
observables are the 1- and 2-body correlators studied in Fig.~\ref{fig:localH}.
\begin{figure}
\centering
\includegraphics[width=5.0in]{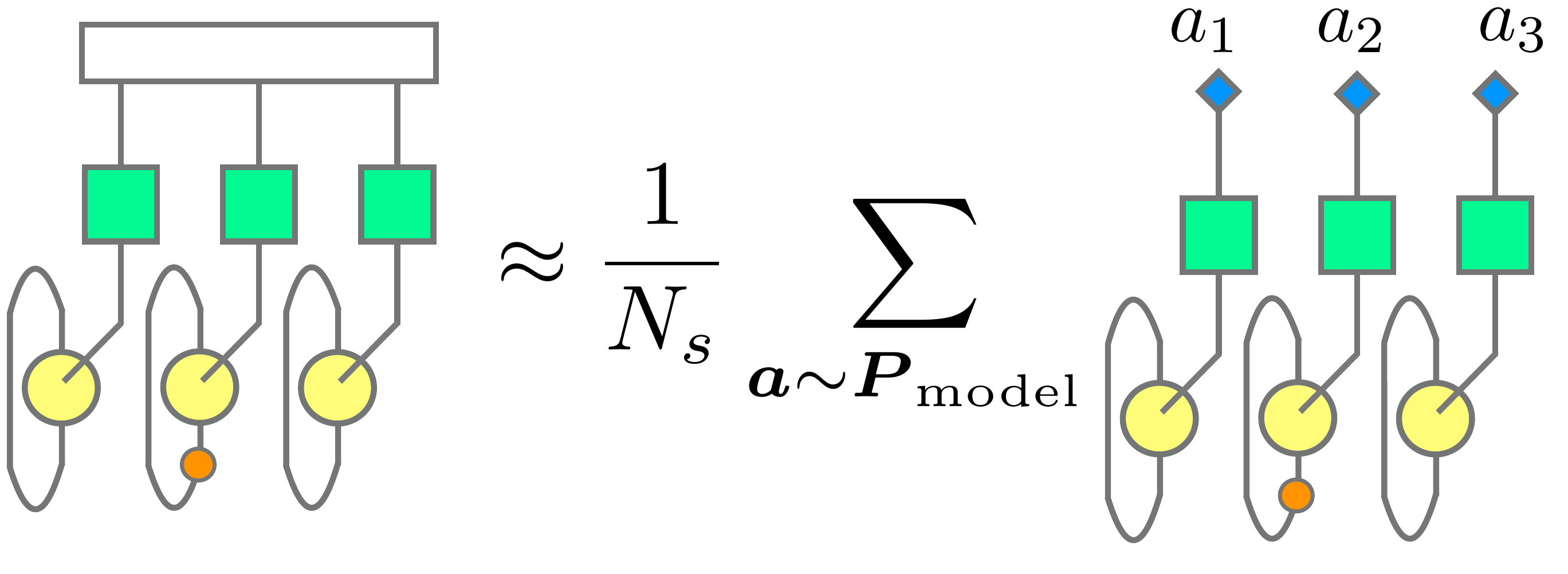}
\caption{
Direct estimation of local observables from $N_s$ model samples (i.e., without $\varrho_{\text{model}}$),
for the exemplary case of a single-qubit observable $O$ (orange tensor) and $N=3$.
The expectation value $\text{Tr}\left[O\,\varrho_{\text{model}}\right]$ (left-hand side)
equals the expected value $\mathbb{E}_{\boldsymbol{a}\sim\boldsymbol{P}_{\text{model}}}\left[Q_{O}(\boldsymbol{a})\right]$
of a random variable $Q_{O}(\boldsymbol{a})$ defined by an $\boldsymbol{a}$-dependent tensor
contraction (right-hand side), with the string $\boldsymbol{a}$ schematically represented by
the blue, rhombus-shaped single-index tensors. The expected value can in turn be efficiently
Monte-Carlo estimated as an average over $N_s$ realizations of
$Q_O(\boldsymbol{a})$ with $\boldsymbol{a}$ sampled from $\boldsymbol{P}_{\text{model}}$.}
\label{fig:local_obs}
\end{figure}
To see this, note first that, since $\boldsymbol{M}$ is IC, one can expand any arbitrary
observable as $O=\sum_{\boldsymbol{a}} Q_O(\boldsymbol{a})\, M^{(\boldsymbol{a})}$,
where $Q_O(\boldsymbol{a})$ are a-priori unknown complex coefficients.
These coefficients are univocally specified by the linear system of equations
$\text{Tr}\left[O\,M^{(\boldsymbol{a}')}\right]=\sum_{\boldsymbol{a}}Q_O(\boldsymbol{a})\,T_{\boldsymbol{a},\boldsymbol{a}'}$, for all $\boldsymbol{a}'$,
which can be straightforwardly solved because $\boldsymbol{T}$ is invertible and factorable.
Then,  using the above expansion, note that $\text{Tr}\left[O\,\varrho_{\text{model}}\right]=\sum_{\boldsymbol{a}}Q_O(\boldsymbol{a})\,P_{\text{model}}(\boldsymbol{a})=\mathbb{E}_{\boldsymbol{a}\sim\boldsymbol{P}_{\text{model}}}\left[Q_{O}(\boldsymbol{a})\right]$.
Now, since $O$ has non-trivial support only a constant number of qubits, the variance of $Q_{O}(\boldsymbol{a})$
(as a random variable over $\boldsymbol{a}$) is independent of $N$. Therefore, the expected value over $\boldsymbol{a}$
sampled from $\boldsymbol{P}_{\text{model}}$ can be estimated efficiently up to arbitrary constant precision
$\varepsilon$ (i.e., with the computational run-time and $N_s$ both scaling polynomially in $N$ and $\varepsilon^{-1}$)
by an average $\frac{1}{N_s}\sum_{\boldsymbol{a}}Q_O(\boldsymbol{a})$ over $N_s$ samples of the random variable
$Q_O(\boldsymbol{a})$. See Fig.~\ref{fig:local_obs}.

Analogously, apart from local observables, one can also directly estimate efficiently the expectation
value of non-local observables $O_{\text{MPO}}$ that admit an efficient matrix-product operator
decomposition, provided that the variance of $Q_{O_{\text{MPO}}}(\boldsymbol{a})$ scales polynomially
with $N$. This includes the quantum fidelity between $\varrho_{\text{model}}$ and a target MPS of
constant bond dimension, again provided that the corresponding variance scales well (see the appendix).

\section*{Data Availability}
The  numerically generated measurements used to produce for Fig.\ref{fig:localH}, the implementation of the generative models, 
as well as code to numerically generate the datasets used in the manuscript are available at 
\url{https://github.com/carrasqu/POVM_GENMODEL}

\section*{Author contributions}
All authors contributed significantly to this work.

\section{Competing interests}
The authors declare no competing interests.


{\textit{Acknowledgments}}.
We would like to thank Guifre Vidal, Lukasz Cincio, and Miles Stoudenmire for discussions and encouragement. We thank
Nathan Berkovits, Alexandre Reily Rocha, and Pedro Vieira for organizing the ICTP-SAIFR/IFT-UNESP
Minicourse on Machine Learning for Many-Body Physics, where this work was started. We thanks Paul Ginsparg for carefully
reading the manuscript and for pointing out several transcription mistakes in earlier versions of the manuscript. 
This research was supported by the Perimeter Institute for Theoretical
Physics, and the Shared Hierarchical Academic
Research Computing Network (SHARCNET). Research at Perimeter Institute is supported through Industry
Canada and by the Province of Ontario through the Ministry of Research \& Innovation. 
R.G.M.~acknowledges support from NSERC of Canada and a Canada Research Chair.
J.C.~acknowledges financial and computational support from the AI grant and Canada CIFAR AI (CCAI) Chairs Program.  
L.A.~acknowledges financial support from the Brazilian agencies CNPq (PQ grant No. 311416/2015-2 and INCT-IQ), FAPERJ (JCN   E-26/202.701/2018), CAPES (PROCAD2013), FAPESP, and the Brazilian Serrapilheira Institute (grant number Serra-1709-17173).

\section*{Supplementary Information}

\section{Generative models}
\subsection{RBM with softmax units}
The number of measurement outcomes at each qubit is $m=6$ for the Pauli-6 and $m=4$ tetrahedral and 
Pauli-4 POVMs. To represent these measurement distributions $\boldsymbol{P}_{\text{model}}$  
for a system with $N$ qubits we first employ an RBM with $N$ $m$-index visible 
units \cite{Salakhutdinov2007}. We encode the states of each unit through one-hot vectors 
with $m$ components on each qubit. A one-hot vector is a $1\times m$ vector used to distinguish 
each measurement outcome. The vector consists of zeros in all cells with the exception of a single 1 
in a cell used uniquely to identify the measurement.  The sampling strategy  we follow uses block 
Gibbs sampling, where groups of two or more variables are sampled together from their joint distribution 
conditioned on all other variables, rather than sampling from each one individually. The conditional 
probabilities are given by
\begin{equation}
p\left(v_i^k =1 \mid \boldsymbol{h} \right)= \frac{e^{ \sum_{j=1}^{n_H} \sum_{l=1}^{L} W_{i j}^{k l} h_{j}^{l}  + b_{i}^k}  }{ \sum_{k'=1}^{m} e^{ \sum_{j}^{n_H } \sum_{l}^{L} W_{i j}^{k'l} h_{j}^{l}  + b_{i}^{k'}} }.
\end{equation}
Here $n_H$ is the number of $L$-dimensional hidden units. $W_{ij}^{kl}$ is the $ijkl$-th element of the weight matrix $W$, 
which is a 4-index array encoding the interaction between the different states $k$ and $l$ 
in the visible and hidden units $i$ and $j$, respectively. 
The $b_i^k$ are the biases in the visible unit and there is an analog expression for the hidden units $a_{j}^l$.  
The energy of the  RBM is given by 
$E(\boldsymbol{v},\boldsymbol{h} ) = -\sum_{ijkl}  W_{ij}^{kl} v_i^k h_j^l - 
\sum_{ik} b_i^k v_i^k -\sum_{ik} a_j^l h_j^l$,
while the joint probability over the visible and hidden units is given by 
$P(\boldsymbol{v},\boldsymbol{h})= \frac{e^{-E(\boldsymbol{v},\boldsymbol{h} )}}{Z}$, where 
$Z= \sum_{ \boldsymbol{v},\boldsymbol{h}} e^{-E(\boldsymbol{v},\boldsymbol{h} )}$.
We train our RBMs using standard block Gibbs sampling and contrastive divergence \cite{Goodfellow-et-al-2016}.

\subsection{Recurrent neural network models}

Recurrent neural networks (RNN) are extensions of the traditional feedforward neural 
networks designed to process sequential data \cite{chung2014empirical,Goodfellow-et-al-2016}. 
In our context, the data corresponds to a measurement outcome string $\boldsymbol{a}=(a_1,a_2,...,a_N)$, 
which we understand as a sequence of single-qubit outcomes according to the given ordering of the qubits. The 
RNNs process sequential data by updating a recurrent hidden state $\boldsymbol{h}_i$ whose 
value at each time is dependent on that of the previous steps, i.e.
\begin{equation}
\boldsymbol{h}_i = \phi\left(\boldsymbol{h}_{i-1}, a_{i}; \boldsymbol{\theta}  \right),
\end{equation}
where $\phi$ is a non-linear function with parameters $\boldsymbol{\theta}$ In its original
form, RNNs process $\boldsymbol{h}_{i-1}$  via 
\begin{equation}
\boldsymbol{h}_i = f\left(W\left[\boldsymbol{a}_{i}; \boldsymbol{h}_{i-1}\right]\right),
\end{equation}
where $\left[\boldsymbol{x};\boldsymbol{y}\right]$ concatenates a k-dimensional vector $\boldsymbol{x}$ and an
q-dimensional vector $\boldsymbol{y}$. 
The matrix $W$ 
contains the trainable parameters of the model. The function $f$ is either a sigmoid $\sigma$ or 
a hyperbolic tangent $\tanh$, both of which act element-wise on any matrix or vector. 
The vector $\boldsymbol{a}_{i}$ 
correspond to a one-hot vector encoding of the integer outcome $a_i$ at qubit $i$.

A generative RNN model outputs a probability distribution over the next element in a 
sequence given all previously observed measurements, i.e.
\begin{equation}
P\left(a_i | a_{1}, \dotsc , a_{i-1} \right) = S\left( \boldsymbol{h}_{i-1}; U\right) 
\end{equation}
where $S$ is a softmax and $U$ is a matrix with parameters optimized during
training. Using the chain rule of probability, the full model is given by
\begin{equation}\label{chainrule}
P_{\text{model}}(a_1, a_2 \dotsc a_{N})= P\left(a_1 \right)P\left(a_2 | a_1 \right) P\left(a_3 | a_1, a_2 \right)  \dotsc P\left(a_N | a_1, a_2 \dotsc a_{N-1} \right). 
\end{equation}

\begin{figure}
\centering
\includegraphics[width=5.5in]{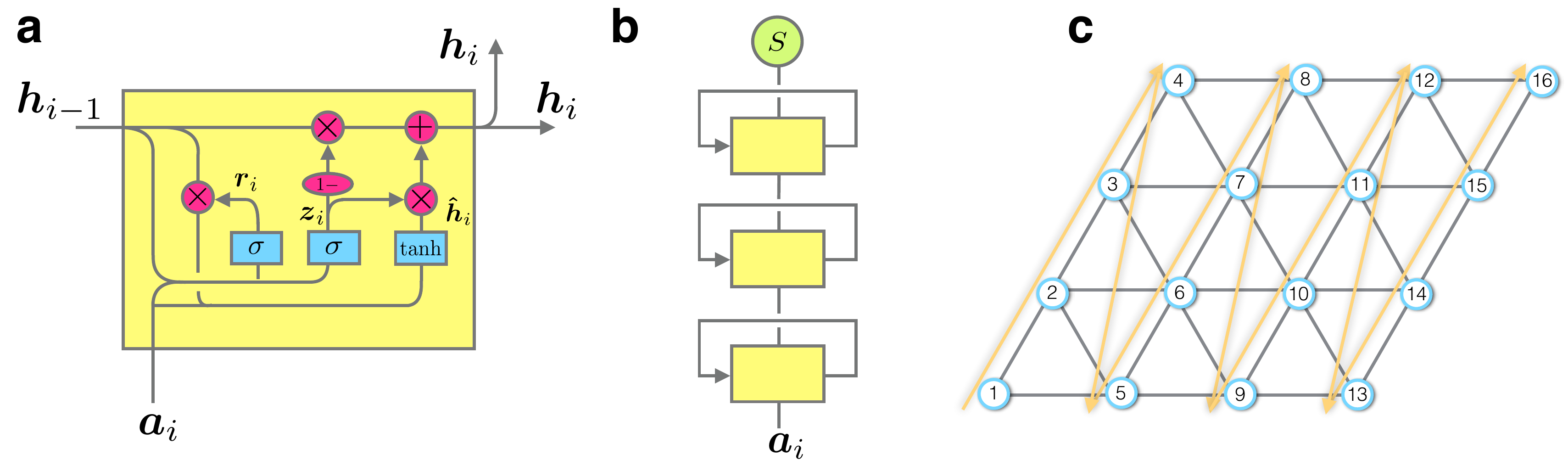}
\caption{
Probabilistic RNN model. (a) Graphical representation of the gated 
recurrent unit described in Eq.~\ref{eq:GRU} and used in our numerical 
experiments. (b) Our deep RNN  model stacks three GRU units (yellow blocks) 
followed by a fully connected layer with a softmax activation denoted by $S$. 
(c) Enumeration and one-dimensional path along which the RNN processes 
the measurement outcomes for the two-dimensional Heisenberg model.}
\label{RNN}
\end{figure}

In its original form, training RNNs is challenging since capturing long-term dependencies between
the qubit measurements tends to make the gradients of the cost function with respect to the parameters
in the RNN either explode or vanish. Furthermore, because of the recurrent structure, 
the long-term dependencies tend to be masked exponentially with respect to the effect of short-term 
dependencies. To overcome these limitations, a possible direction is to modify the recurrent unit 
using a long short-term memory (LSTM) unit \cite{hochreiter1997long}, which are explicitly developed 
to avoid the long-term dependency problem. Similarly, the gated recurrent unit (GRU) \cite{W14-4012}, which 
was originally introduced in the context of neural machine translation, is designed to adaptively 
capture dependencies of different scales. The GRU processes the sequential data through
\begin{align}
\label{eq:GRU}
\boldsymbol{z}_i &=\sigma\left( W_z\left[ \boldsymbol{h}_{i-1};\boldsymbol{a}_{i}\right]\right)  \\
\nonumber
\boldsymbol{r}_i &=\sigma\left( W_r\left[ \boldsymbol{h}_{i-1};\boldsymbol{a}_{i}\right]\right)  \\
\boldsymbol{\hat{h}}_i&= \tanh\left( W_c\left[ \boldsymbol{r}_i\odot \boldsymbol{h}_{i-1}  ;\boldsymbol{a}_{i} \right] \right).\\
\boldsymbol{h}_{i} & = (1-\boldsymbol{z}_i)\odot\boldsymbol{h}_{i-1} + \boldsymbol{z}_i \odot \boldsymbol{\hat{h}}. 
\end{align}
Here the hidden state $\boldsymbol{h}_{i}$ is updated through a linear interpolation between 
the previous activation $\boldsymbol{h}_{i-1}$ and the candidate hidden state $\boldsymbol{\hat{h}}_i$.
The update gate $\boldsymbol{z}_i$ decides how much to update the contents of the hidden state. 
Here $\boldsymbol{r}_i$ is a set of reset gates and $\odot$ denotes  element-wise vector multiplication.  
When the reset gates are ``off'', i.e. $\boldsymbol{r}_i\approx 0$, they cancel out the unit which then acts 
as if it is reading the first element of the sequence, effectively making the unit ``forget'' part of 
the sequence that has already been encoded in the state vector $\boldsymbol{h}_{i}$, if necessary. 
The matrices $W_{z,r,c}$ parametrize the GRU and are optimized using standard
maximum likelihood estimation. The GRU in Eq.~\ref{eq:GRU} are graphically represented in Fig.~\ref{RNN}.  
For Fig.~\ref{RNN}, we emphasize that the meaning of lines, circles, and squares is 
different from the tensor-network notation used througout the text. Here, lines with 
arrowheads denote incoming vectors from the output of one node to the inputs of others. 
The magenta circles/elipses represent pointwise operations such as vector addition or multiplication.  
The blue rectangles represent neural network layers labeled by the type of nonlinearity they used. 
Lines merging denote vector concatenation, while lines forking means the vector is copied and 
used in another operation.  

To make our models more expressive we train a deep RNN model with three stacked 
GRU units (see Fig.~\ref{RNN}(b)). In all of our experiments the dimension of 
the state vector $\boldsymbol{h}_{i}$ is set to 100.  In one spatial dimension 
there is a natural enumeration scheme for the lattice sites (e.g. from right to 
left or viceversa) along which the RNN model processes the data. In the absence 
of a notion of spatial dimensionality, like in the GHZ state, one can choose an 
arbitrary enumeration of the qubits. In two dimensional systems, such as the 
triangular Heisenberg model in the manuscript, we choose an enumeration of the 
lattice sites which, in turn, defines a one-dimensional path filling the 
two-dimensional lattice. The RNN then uses this path to process and learn the data. This is 
exemplified in Fig.~\ref{RNN}(c) for a $4\times4$ triangular lattice. An analogous 
enumeration is used in our experiments in the main text. 

\section{Direct stochastic estimation of quantum fidelity with respect to an MPS}
\begin{figure}
\centering
\includegraphics[width=5.0in]{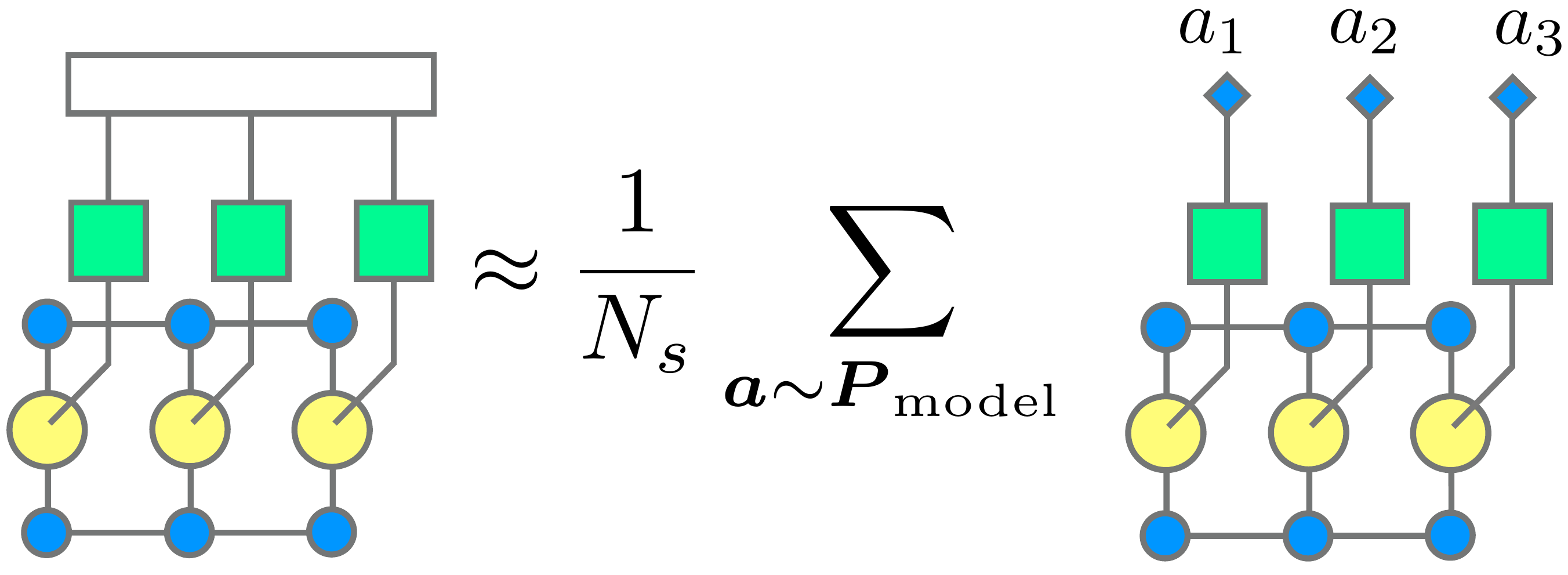}
\caption{
Estimation of the quantum fidelity between $\varrho_{\text{model}}$ and a target
$\ket{\Psi}$ directly from $N_s$ samples from the model distribution
$\boldsymbol{P}_{\text{model}}$ for $N=3$. If $\ket{\Psi}$ is an MPS (blue circles)
of constant bond dimension, it can be expressed as the virtual-index
contraction of 2- and 3-index tensors of constant size. The squared fidelity
corresponds to the contraction of the MPS with $\varrho_{\text{model}}$ (left-hand side).
This, in turn,  can  be efficiently Monte-Carlo estimated as an average
(right-hand side) over $N_s$ realizations of $Q_{\ketbra{\Psi}{\Psi}}(\boldsymbol{a})$
with $\boldsymbol{a}$ sampled from $\boldsymbol{P}_{\text{model}}$}.
\label{fig:fid_MPS}
\end{figure}

For POVMs with invertible overlap matrix, the quantum fidelity between a reconstructed 
state $\varrho_{\text{model}}$ and a generic pure  state $\ket{\Psi}$ can in principle 
be computed sampling the model distribution $\boldsymbol{P}_{\text{model}}$. Since one 
of the states is pure, the squared fidelity is given by 
\begin{equation}
F^2\left(\ketbra{\Psi}{\Psi},\varrho_{\text{model}}\right)=\mathbb{E}_{\boldsymbol{P}_{\text{model}}}
\left[Q_{\ketbra{\Psi}{\Psi}}(\boldsymbol{a}) \right],
\end{equation}
with $Q_{\ketbra{\Psi}{\Psi}}(\boldsymbol{a})=\sum_{\boldsymbol{a}'} 
T^{-1}_{\boldsymbol{a},\boldsymbol{a}'}\bra{\Psi} M^{(\boldsymbol{a}')}\ket{\Psi}$.
Whereas the computation of each $Q_{\ketbra{\Psi}{\Psi}}(\boldsymbol{a})$ 
is in general intractable, it can be done efficiently for any state $\ket{\Psi}$ admitting 
an efficient MPS representation (see Fig.~\ref{fig:fid_MPS}). In addition, however, for 
the Monte-Carlo estimation to converge efficiently, it is also required that 
the variance of the estimtor $Q_{\ketbra{\Psi}{\Psi}}(\boldsymbol{a})$  does not grow 
too fast in $N$. Unfortunately, we find that the variance of $Q_{\ketbra{\Psi}{\Psi}}(\boldsymbol{a})$ 
for the $N$-qubit GHZ state (see Sec. \ref{sec:GHZ_MPS} for its MPS representation) 
grows dramatically with $N$, making the estimation unfeasible for systems as small as $N\approx 8$. 
It is an open question whether there are non-trivial MPS states for which the 
corresponding variance grows slowly with $N$.

\section{Matrix product operator representation of the noisy GHZ and its sampling}
\label{sec:GHZ_MPS}
\begin{figure}
\centering
\includegraphics[width=4.5in]{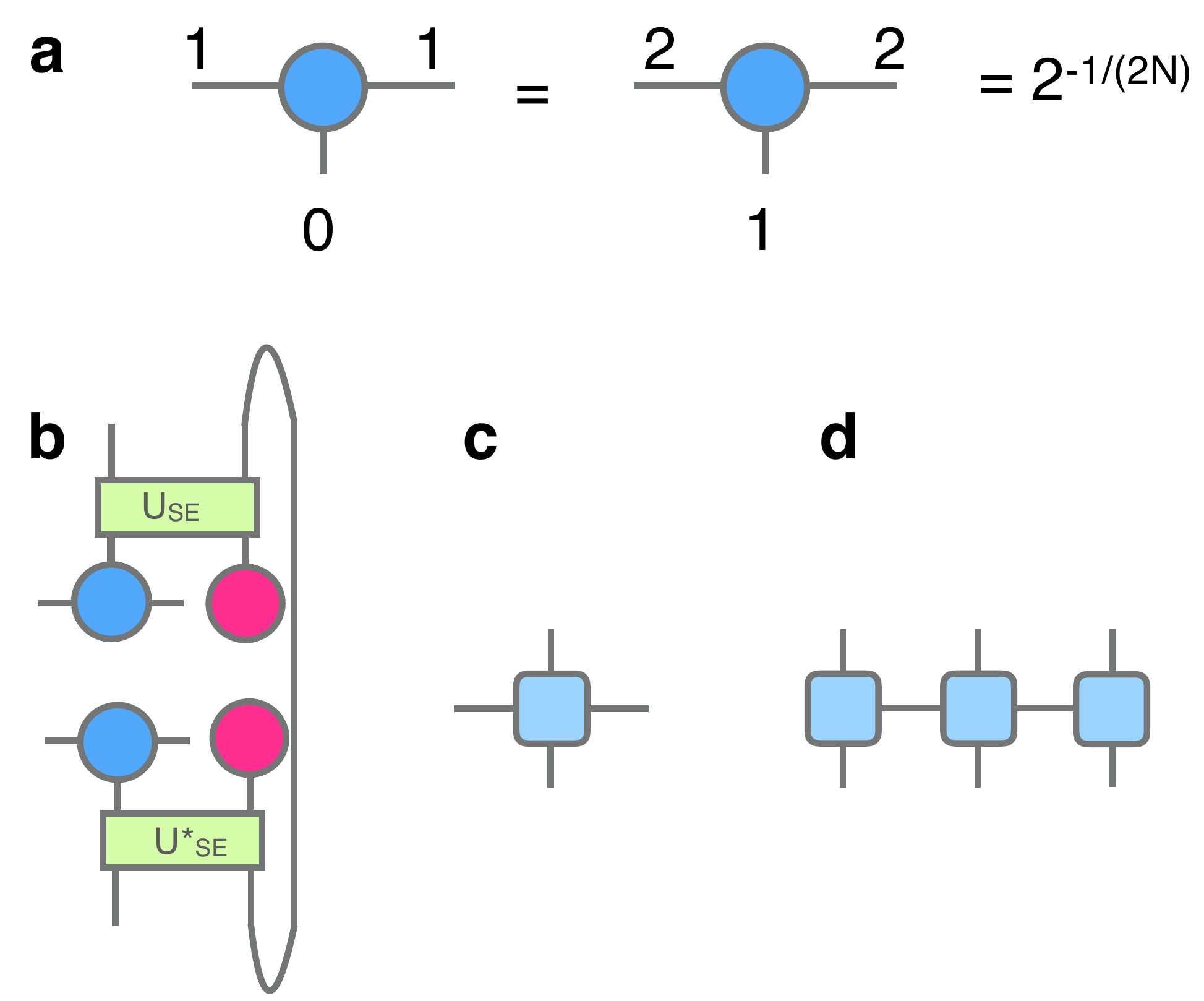}
\caption{Tensor-network description of the GHZ states. (a)The non-zero elements of the tensors
representing the GHZ state as an MPS 
(b) Applying the unitary operator $U_{\text{SA}}$ locally
to the tensors in the pure GHZ and their ancillae. (c) The tensor in (b) is reshaped to a $4$ dimensional
tensor making up the MPO representation of the noisy GHZ state in (d).}
\label{GHZ_MPO}
\end{figure}

The GHZ state in the absence of noise $p=0$ can be written down as an MPS of bond dimension
$\chi=2$~\cite{Orus:2013kga}. The non-zero elements of the MPS tensors are displayed in 
Fig.~\ref{GHZ_MPO} (a). Similarly, the locally depolarized GHZ state can be expressed as an
MPO with bond dimension $\chi=4$ as follows. 
Given an $N$-qubit MPS representation of the pure GHZ, attach an ancilary ququart 
initialized in a reference state $|0\rangle_A$ to each of the $N$ system qubits. Then 
apply a local unitary operator $U_{\text{SA}}$ acting on the system qubit (S) and ancilla
(A), which realizes a unitary dilation of the depolarizing channel. Once the ancilla is 
traced out, the desired locally depolarized state is obtained on S. More precisely,  the single-qubit
depolarizing channel can be realized by an isometry mapping the state of the qubit $|\Psi \rangle_A$ 
to (qubit + ancilla ququart) $| \Psi\rangle_{AE}$ state acting as \cite{PreskillNotes} 
\begin{equation}
\label{eq:Unitary}
|\Psi \rangle_A  \longmapsto  | \Psi\rangle_{AE}= \sqrt{1-p} |\Psi \rangle_S \otimes |0\rangle_{A} 
+ \sqrt{\frac{p}{3}} \left( \sigma^{x}_S |\Psi \rangle_S \otimes |1 \rangle_{A} 
+ \sigma^{y}_S |\Psi \rangle_S \otimes |2 \rangle_{A}
+ \sigma^{z}_S |\Psi \rangle_S \otimes |3 \rangle_{A}\right),
\end{equation} 
and then subsequently tracing the ancila out (see 
Fig. \ref{GHZ_MPO} (b)). On an $N$-qubit system, apply this operation on each of the qubits. 
The resulting tensor is reinterpreted and reshaped 
as the factors (Fig. \ref{GHZ_MPO} (c)) in an MPO  that represents the density matrix 
of the system, as depicted Fig. \ref{GHZ_MPO} (d). 

\begin{figure}
\centering
\includegraphics[width=4.5in]{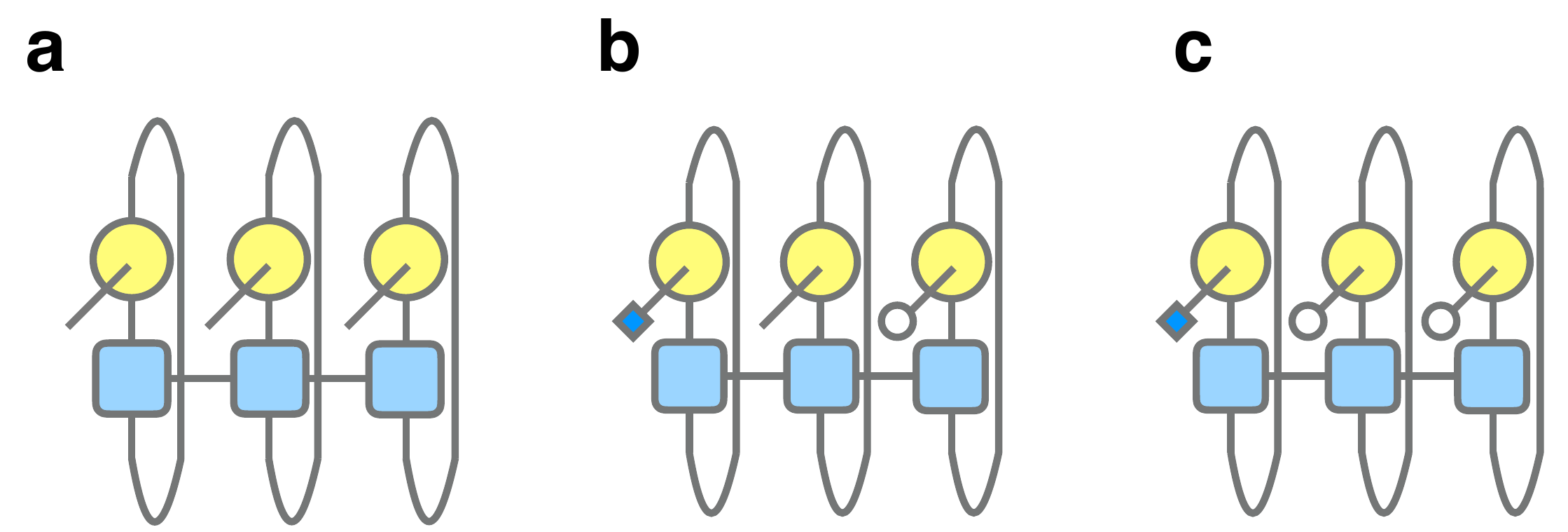}
\caption{Sampling an outcome string $\boldsymbol{a}$ from $\boldsymbol{P}_{\text{MPO}}$. 
(a) Tensor-network representation of $\boldsymbol{P}_{\text{MPO}}$.
The marginal conditional probability $P\left(a_2|a_1\right)$ is calculated as the ratio 
of the vector (b) and normalization constant (c).} 
\label{Sampling_MPO}
\end{figure}

Given a factorable POVM, and an MPO  density matrix 
with small bond dimension $\chi$, the measurement statistics $\boldsymbol{P}_{\text{MPO}}$, 
depicted graphically in Fig. \ref{Sampling_MPO} (a), 
can be tractably sampled using the ideas introduced in Ref. \cite{PhysRevB.85.165146}. 
Our strategy invokes the chain rule of probability in Eq. \ref{chainrule} and computes
sequentially a series of conditional single-qubit probabilities $\{P\left(a_1\right), 
P\left(a_2|a_1\right), \hdots  P(a_N|a_{i< N}) \}$, each of which can be computed efficiently for the 
tensor network representation in Fig. \ref{Sampling_MPO} (a). First we compute
$P(a_1)$ by  contracting the entire tensor network from right to left with
constant vectors (white vectors $\boldsymbol{e}=(1,1,1,1)$ in Fig. \ref{Sampling_MPO} (b)). 
Then we sample $P(a_1)$ and fix its index to the obtained sample $a_1'$ [blue diamond vector 
with components $a^*_{i}=\delta_{i,a_1'}$ in Fig. \ref{Sampling_MPO} (b)]. 
Proceed to site 2 and compute $P\left(a_2|a_1\right)$ which is the ratio between 
$P\left(a_1',a_2\right)=\sum_{a_3,\hdots, a_N} P\left(a_1', a_2, a_3,\hdots, a_N \right)$ 
[graphically depicted in Fig. \ref{Sampling_MPO} (b)] and 
$P(a_1')=\sum_{a_2,a_3,\hdots, a_N} P\left(a_1', a_2, a_3,\hdots, a_N \right)$ presented in 
Fig. \ref{Sampling_MPO} (c). Sample $P\left(a_2|a_1\right)$ and continue doing the same 
for the rest of qubits in the system until the $N$-th qubit is reached. Note that apart 
from the exact sample $\boldsymbol{a}'=(a_1', a_2', a_3',\hdots, a_N')$, its probability 
is also tractable and given by Eq. \ref{chainrule}. 
 
\section{Density-matrix renormalization group calculations}
The training datasets for the reconstruction of the ground states of spin Hamiltonians have 
been generated using the density matrix renormalization group (DMRG). The algorithm is 
implemented within the framework of MPSs, and it is executed using the ITensor 
library~\cite{ITensor}. Given a particular spin model, with the Hamiltonian expressed as a 
MPO, the DMRG algorithm attempts to find the optimal MPS with the lowest energy. The elements 
of the $N$ tensors in the MPS are optimized using a unit cell containing two sites. For the 
one-dimensional antiferromagnetic Ising model at the critical point, the ground state is obtained 
by performing 5 DMRG sweeps along the chain. The resulting MPS for $N=50$ spins has bond dimension 
$\chi=25$, and a cut-off error of $9.93\times10^{-10}$. The second spin model we considered is 
the Heisenberg model on a two-dimensional triangular lattice. Given the symmetries of the Hamiltonian, 
we restrict the DMRG calculations to the sector corresponding to a zero total magnetization, 
$\sum_i\sigma^z_i=0$. To find an approximation to the ground state of the model, we perform 
10 DMRG sweeps with a maximum bond dimension of $\chi=200$, and a resulting truncation error of $1.8\times10^{-4}$.

\end{document}